\documentclass{PoS}

\title{Bottomonium and B results from full lattice QCD}

\ShortTitle{Bottomonium and B results from full lattice QCD}

\author{HPQCD Collaboration}

\author{\speaker{C. T. H. Davies}, B. Colquhoun, B. Galloway, G. C. Donald, J. Koponen\\
        SUPA, School of Physics and Astronomy, University of Glasgow, Glasgow, G12 8QQ, UK\\
        E-mail: \email{Christine.Davies@glasgow.ac.uk}}

\author{R. J. Dowdall, R. Horgan\\
        DAMTP, University of Cambridge, Wilberforce Road, Cambridge, CB3 0WA, UK}

\author{E. Follana\\
         Departamento de F\'{i}sica Te\'{o}rica, Universidad de Zaragoza, E-50009 Zaragoza, Spain  }

\author{G. P. Lepage\\
         Labortory for Elementary-Particle Physics, Cornell University, Ithaca, NY 14853, USA  }

\author{C. McNeile\\
        School of Computing and Mathematics, Plymouth University, Drake Circus, Plymouth, PL4 8AA, UK   }

\abstract{ We have developed two methods for handling $b$ quarks 
in lattice QCD. One uses NRQCD (now improved to include radiative 
corrections) and the other uses Highly Improved Staggered Quarks (HISQ), 
extrapolating to the $b$ quark from lighter masses and using multiple lattice spacings to control discretisation errors.
Comparison of results for the two different methods gives confidence 
in estimates of lattice QCD systematic errors, since they 
are very different in these two cases. Here we show results for 
heavyonium hyperfine splittings and vector current-current correlator moments using 
HISQ quarks, to add to earlier results testing the heavy HISQ method with 
pseudoscalar mesons. We also show the form factor 
for $B \rightarrow \pi l \nu$ decay 
at zero recoil using NRQCD $b$ quarks and $u/d$ quarks with physical masses. 
This allows us to test the soft pion theorem relation ($f_0(q^2_{max})=f_B/f_{\pi}$) 
accurately and we find good agreement as $M_{\pi} \rightarrow 0$. } 

\FullConference{31st International Symposium on Lattice Field Theory - LATTICE 2013\\
		July 29 - August 3, 2013\\
		Mainz, Germany}

\begin{document}

\section{Vector Heavyonium - hyperfine splitting}
\begin{figure}
\centering
\includegraphics[width=0.6\textwidth]{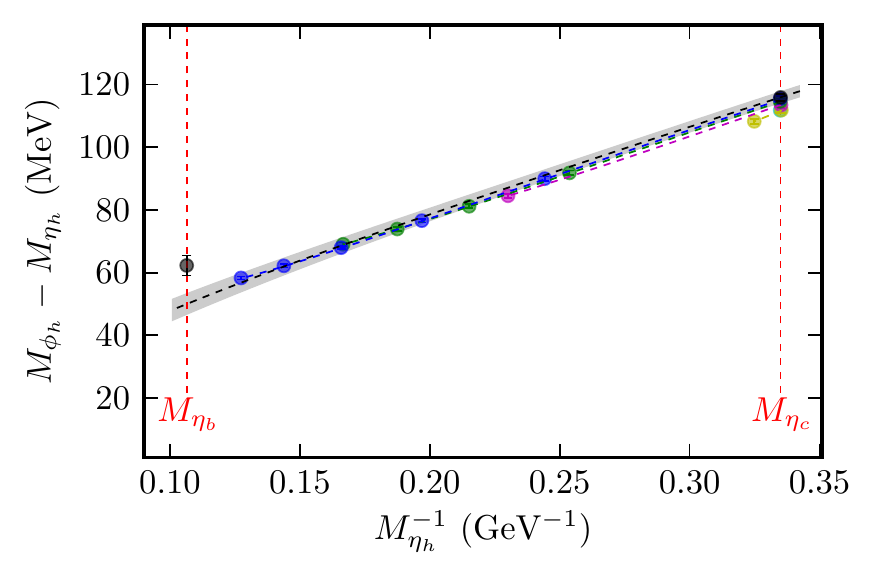}
\caption{The heavyonium hyperfine splitting as a function 
of the inverse heavyonium mass. Results are given in different 
colours for 
coarse ($a$=0.12fm), fine ($a$=0.09fm), superfine ($a$=0.06fm) 
and ultrafine ($a$=0.045fm) 
lattices, including 2+1 flavours of sea quarks. 
The shaded band shows the fit described in the text and
the black points at the $b$ and $c$ the current experimental 
averages~\cite{pdg}. Note that $\eta_b$ and $\eta_c$ annihilation 
effects are \emph{not} included in the lattice results or the shaded fit. } 
\label{fig:hyp}
\end{figure}

Fig.~\ref{fig:hyp} shows the heavyonium vector-pseudoscalar 
(hyperfine) mass splitting as a function of inverse 
heavyonium mass for HISQ quarks~\cite{hisq} for a range of masses
from charm upwards. We have used MILC gluon 
field configurations that include 2+1 flavours of 
asqtad sea quarks with lattice spacings ranging from 
0.12 fm to 0.045 fm~\cite{milcasqtad}. The finer lattice spacings allow 
a much larger reach in meson mass because a given meson 
mass corresponds to a smaller quark mass in lattice units. 
We have used quark masses in lattice units up to 0.8. 
Having so many values of the lattice spacing allows the 
discretisation errors to be well determined from the fit. 
The grey band shows the result of the physical heavy 
quark mass dependence determined at zero lattice spacing resulting 
from a fit function 
of form~\cite{craig}: 
\begin{equation}
F(M,a) = A \left(\frac{M}{M_0}\right)^b \sum_{i=0}^7\sum_{j=0}^3 c_{ij} \left(\frac{M_0}{M}\right)^i(am)^{2j} 
\label{eq:fit}
\end{equation}  
where $M$ is the pseudoscalar meson mass and $m$ the 
quark mass and we also include terms allowing for sea 
quark mass dependence (which are then extrapolated to the 
physical point). Priors on the coefficients are 
generally taken as $0(1)$ and $M_0$ is taken as 1 GeV. 

The result at the $b$, after an increase of 3(3) MeV 
for $\eta_b$ annihilation effects not included in 
our calculation (and also NOT included in the plot) is 
\begin{equation}
M_{\Upsilon}-M_{\eta_b} = 53(5) \mathrm{MeV}. 
\end{equation}
This is in reasonable agreement with the current 
experimental average of 62.3(3.2) MeV~\cite{pdg}. 
The result we obtain at $c$ is 116.5(3.2) MeV~\cite{gordonpsi}, 
in good agreement with the experimental average of 
113.2(7) MeV~\cite{pdg}.  
We are currently extending these calculations to the 
MILC `second-generation' configurations that include 
2+1+1 flavours of HISQ sea quarks~\cite{milchisq}. 

We have also calculated the bottomonium hyperfine splitting 
using NRQCD for the $b$ quark~\cite{rachelhyp}. 
We have included spin-dependent relativistic corrections through 
$\cal{O}$$(v^6)$, radiative corrections to the leading 
spin-chromomagnetic coupling (at $\cal{O}$$(v^4)$) and, for 
the first time, non-perturbative 4-quark interactions. 
We use the
MILC configurations that include 
2+1+1 flavours of HISQ sea quarks. 
We obtain a splitting of 62.8(6.7) MeV~\cite{rachelhyp}, again in good 
agreement with experiment. 

\section{Vector Heavyonium - current-current correlator moments}
\begin{figure}
\centering
\includegraphics[width=0.6\textwidth]{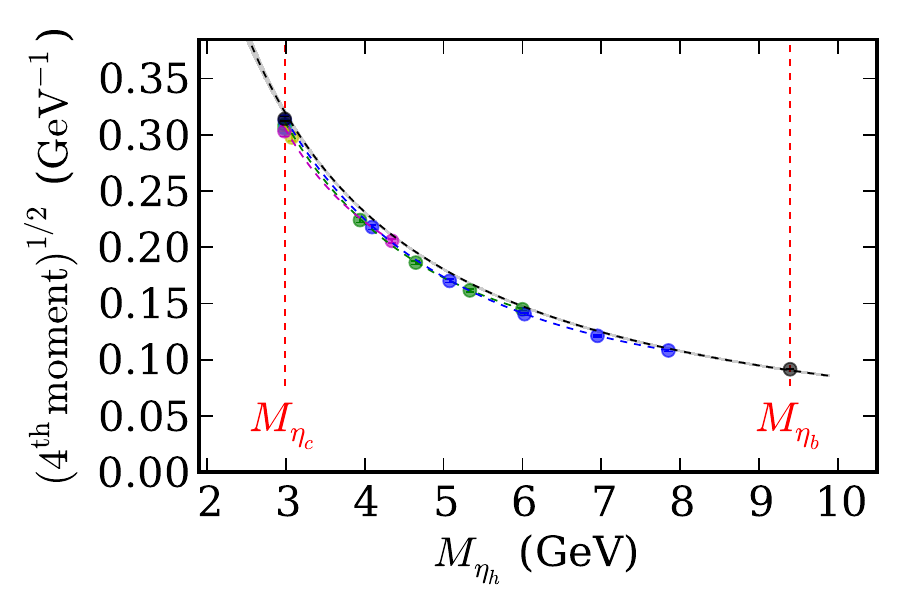}\\
\includegraphics[width=0.6\textwidth]{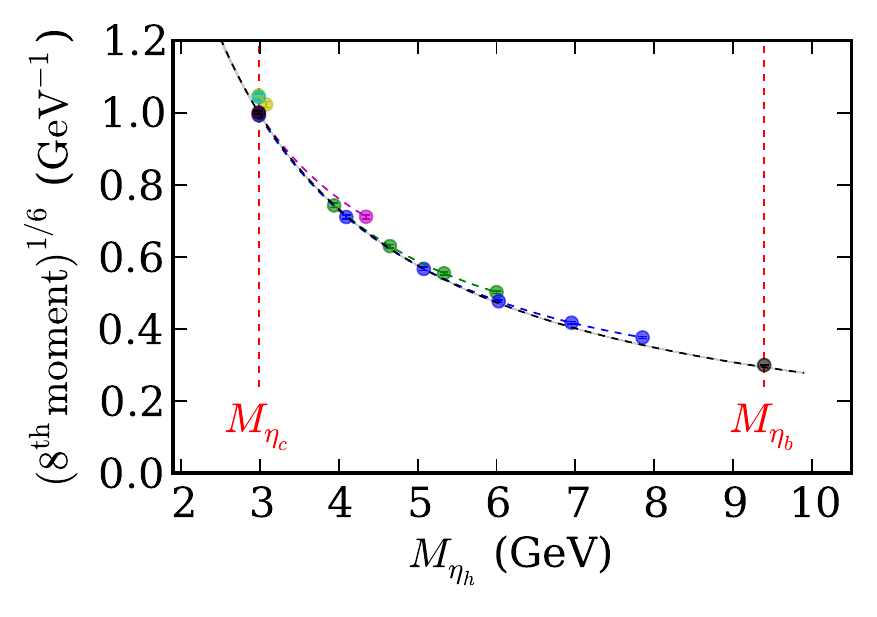}
\caption{The 4th and 8th moments of the heavyonium 
vector current-current correlator as a function 
of heavyonium mass. 
Results are given for 
coarse ($a$=0.12fm), fine ($a$=0.09fm), superfine ($a$=0.06fm) 
and ultrafine ($a$=0.045fm) 
lattices, including 2+1 flavours of sea quarks. 
The shaded band shows the fit described in the text and
the black points at the $b$ and $c$ are the results 
derived from experiment~\cite{chetyrkin}. }
\label{fig:moments}
\end{figure}

Time moments of the vector heavyonium correlator are defined by:
\begin{equation}
G^V_n = Z^2 \sum_{\tilde{t}}\tilde{t}^n C_V(\tilde{t})
\end{equation}
where $\tilde{t}$ is the lattice time variable, symmetrised 
around the centre of the lattice and $Z$
is the current renormalisation factor, derived from 
continuum QCD perturbation theory for the 6th moment~\cite{gordonpsi}. 

The lattice time-moments can be compared to 
$q^2$-derivative moments of the heavy quark vacuum polarisation. 
Values for these can be extracted~\cite{chetyrkin} from experimental 
results in the $c$ and $b$ regions for $R_{e^+e^-}$ where 
\begin{equation}
R_{e^+e^-} = \frac{\sigma(e^+e^- \rightarrow \mathrm{hadrons})}{\sigma_{\mathrm{point}}} .
\end{equation}

The two plots in Fig.~\ref{fig:moments} show our 
results for the 4th and 8th moments, calculated 
with HISQ quarks, as a function of heavyonium 
mass. Again we have results for a wide range of 
lattice spacing values, from 0.12 fm to 0.045 fm, 
on the MILC gluon field configurations that 
include 2+1 flavours of asqtad sea quarks.   

The shaded bands show the results of fits to 
the form given above for the hyperfine splitting (Eq.~\ref{eq:fit})
but with leading power, $b$=-1. 
The results show excellent agreement with experiment, 
both at $c$~\cite{gordonpsi} and at $b$. The experimental 
values~\cite{chetyrkin} are shown as the black points. 
This is a stringent test of lattice QCD since experimental 
errors are small. 

\section{$B \rightarrow \pi l\nu$ decay and soft pion theorems}
\begin{figure}
\centering
\includegraphics[width=0.6\textwidth]{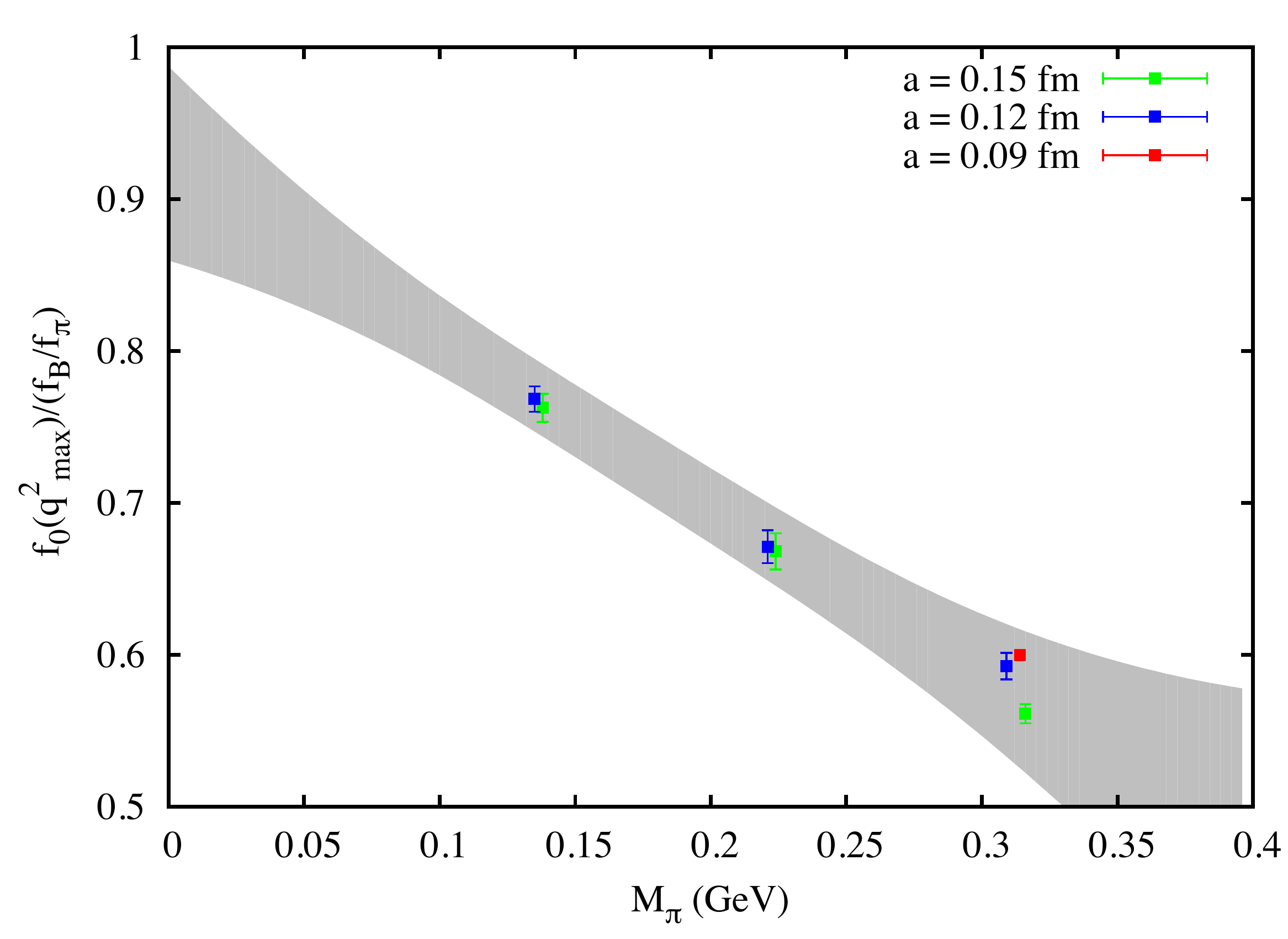}
\caption{The ratio of $f_0(q^2_{max})$ for $B \rightarrow \pi l \nu$ 
decay to $f_B/f_{\pi}$ as a function of $M_{\pi}$. The points 
are lattice results, at 3 values of the lattice spacing, 
with the leftmost points being at the 
physical value of $M_{\pi}$. The shaded band shows a simple extrapolation 
through the two most chiral sets of points to $M_{\pi}=0$ and $a=0$.} 
\label{fig:f0}
\end{figure}

The matrix element of the temporal vector current for 
$B \rightarrow \pi$ decay at zero recoil (both mesons 
at rest) is given by: 
\begin{equation}
\langle \pi | V^4 | B \rangle = f_0(q^2_{max})(M_B+M_{\pi}) .
\end{equation}
Soft pion theorems relate this to decay constants. At leading 
order, for $M_{\pi} \rightarrow 0$, 
\begin{equation}
f_0(q^2_{max})=\frac{f_B}{f_{\pi}} .
\end{equation}
This result seemed not to hold well in the 
quenched approximation (see~\cite{hashimoto} for a review), 
but large uncertainties arose 
from large pion masses (along with the absence of significant 
pion mass dependence), quenching and current renormalisation. 

Here we have small uncertainties because we are using 
improved NRQCD $b$ quarks combined with HISQ light 
quarks. We go down to physical pion 
masses on the MILC configurations that include 
2+1+1 flavours of HISQ sea quarks. $Z$ factors cancel 
between $f_0$ and $f_B$ from staggered chiral symmetry. 
We include $\cal{O}$$(\Lambda/m_b)$ relativistic corrections to 
axial and vector currents. 

The plot in Fig.~\ref{fig:f0} shows the ratio of $f_0$ to the 
decay constant ratio as a function of $M_{\pi}$. 
The decay constants are the current lattice state-of-the-art 
results obtained in~\cite{rachelfb} and~\cite{rachelfpi}. 
We see the ratio is around 0.75, quite far from 1, for 
physical $\pi$ masses. However, there is clearly strong dependence 
on $M_{\pi}$. Using an extrapolation form 
that includes linear terms in $M_{\pi}$ (coming from 
the dependence of $q^2_{max}$ on $M_{\pi}$~\cite{ukqcd}) readily 
gives a result in agreement with 1 in the $M_{\pi} \rightarrow 0$ limit. 

We are currently calculating form factors for $B/B_s$ semileptonic decays 
using HISQ quarks and extrapolating to the $b$ as described above. This 
will give us another opportunity to test systematic errors by comparing 
the two formalisms.

\end{document}